\documentclass[10pt,journal,compsoc]{IEEEtran}

\usepackage[ruled]{algorithm2e} 

\SetAlFnt{\small}
\SetAlCapFnt{\small}
\SetAlCapNameFnt{\small}
\SetAlCapHSkip{0pt}
\IncMargin{-\parindent}

\usepackage{soul}
\soulregister\cite7
\soulregister\ref7

\usepackage{listings}
\usepackage{color}
\usepackage{amsthm}
\usepackage{amsmath}
\usepackage{graphicx}
\usepackage[table]{xcolor}
\usepackage[bookmarks=false]{hyperref}
\usepackage{booktabs}
\usepackage[normalem]{ulem}
\useunder{\uline}{\ul}{}
\def\BibTeX{{\rm B\kern-.05em{\sc i\kern-.025em b}\kern-.08em
    T\kern-.1667em\lower.7ex\hbox{E}\kern-.125emX}}
\usepackage[noadjust]{cite}
\usepackage{multirow}
\usepackage{rotating}
\usepackage{caption}

\definecolor{dkgreen}{rgb}{0,0.6,0}
\definecolor{gray}{rgb}{0.5,0.5,0.5}
\definecolor{mauve}{rgb}{0.58,0,0.82}

\begin{document}

\title{Satisfying Increasing Performance Requirements with Caching at the \\ Application Level}

\author{Jhonny~Mertz,
        Ingrid~Nunes,
        Luca~Della~Toffola,
        Marija~Selakovic
        and~Michael~Pradel
\IEEEcompsocitemizethanks{\IEEEcompsocthanksitem J. Mertz and I. Nunes are with Universidade Federal do Rio Grande do Sul, Porto Alegre, Brazil.
\IEEEcompsocthanksitem L. Della Toffola was with ETH Zurich, Zurich, Switzerland.
\IEEEcompsocthanksitem M. Selakovic and M. Pradel are with TU Darmstadt, Darmstadt, Germany.}
}

\IEEEtitleabstractindextext{%
\begin{abstract}
Application-level caching is a form of caching that has been increasingly adopted to satisfy performance and throughput requirements. The key idea is to store the results of a computation, to improve performance by reusing instead of recomputing those results. However, despite its provided gains, this form of caching imposes new design, implementation and maintenance challenges. In this article, we provide an overview of application-level caching, highlighting its benefits as well as the challenges and the issues to adopt it. We introduce three kinds of existing support that have been proposed, giving a broad view of research in the area. Finally, we present important open challenges that remain unaddressed, hoping to inspire future work on addressing them.

\textbf{Key Insights:}
\begin{itemize}
\item Application-level caching saves precious computing cycles by reusing already computed values.
\item Application-level caching provides significant performance gains but is non-trivial to implement and to maintain.
\item A variety of research findings, tools, and methods are available for using application-level caching effectively.
\item Automating the detection and exploitation of caching opportunities can help developers to speed up their code without sacrificing maintainability.
\end{itemize}
\end{abstract}

\begin{IEEEkeywords}
application-level caching, performance, cache, memoization, software development.
\end{IEEEkeywords}}

\maketitle

\IEEEdisplaynontitleabstractindextext
\IEEEpeerreviewmaketitle

\section{Introduction}

Web applications pushed performance requirements to new limits. In a single second, over 73k Google searches are made, and over 78k YouTube videos are viewed. To support these intensive workloads, software engineers are required to design and implement scalable solutions backed by a large computational infrastructure. A common approach to scale an application to larger workloads is to use caching. Caching is an effective means of improving software performance by storing the content that is accessed in locations that can be reached faster than the original location.

Caching can be introduced at different levels. CPUs, for instance, make use of fast hardware caches to speed-up computations by storing frequently accessed memory locations. Similarly, caching can be introduced in the network (e.g., as a proxy server) or within applications (e.g., to store recent accesses to a database). These different cache components are an important part of the infrastructure of many major Internet companies, including Zynga, Twitter, Tumblr, Reddit, Facebook, and Wikipedia. Twitter, for example, has 5.9\% of its hardware and 7.3\% of its storage dedicated for caching~\cite{twitter:2017}.

Due to the increasing demand for computational resources to continue satisfying users' requirements, software developers are challenged to continuously improve performance and throughput of applications. A popular way to achieve it is by caching the results of any computation that produces content likely to improve performance if reused instead of recomputed. This kind of caching---referred to as \emph{memoization} or \emph{application-level caching}, the latter being the term we use in this article---has been leveraged to deal with the significant amount of user-tailored content that can be generated by applications nowadays.

Application-level caching is used for two primary purposes: as in-memory storage to reduce the load on data source providers and as a buffer to store items that are expensive to recompute~\cite{Mertz2017}. A study~\cite{Selakovic2016} showed that 13\% of all performance issues in JavaScript-based applications are caused by repeated operations, which could have been avoided with the use of application-level caching.

To benefit from application-level caching, developers manually identify content that can be cached and then modify the target application to do so. This implies that the design and implementation of application-level caching are done in an ad-hoc manner for each specific application. As a result, although promising to improve application performance, this form of caching comes with a high maintenance cost. In addition, performance becomes a priority only when the application starts struggling with increased access demands. The combination of manual effort and lack of appropriated time creates a situation in which developers may not afford a well-designed cache, leaving room for technical debts and misuses. As a consequence, application-level caching has been commonly underused~\cite{underused:2015,underused:2017,Mertz2017,Selakovic2016}, preventing the maximum exploitation of this powerful optimisation technique.

This article thus raises awareness of the key benefits and costs associated with application-level caching. This caching technique is different from traditional ones---such as HTTP-based, proxy-level, search engines or database caching---because it is not transparent to developers and is placed within the application code. We provide an overview of the tasks to be performed to develop an application-level cache and practical solutions to support it. To encourage the development of future approaches, open challenges associated with this type of caching are also introduced. 


\section{Application-level Caching}

Application-level caching is characterized as caching implemented within the application boundaries, targeting operations of, e.g.\ business and persistence logic. Cached content enables the reuse of previous computations with the goal of decreasing the processing time~\cite{Mertz2018a}. Recent studies demonstrated that application-level caching can yield significant performance and throughput improvements~\cite{Mertz2017,Selakovic2016,Chen2016}. To illustrate how these improvements can be achieved by introducing caching at the application level, Listing~\ref{lst:snippet1} gives an example of a typical caching implementation. In the example, a caching component \texttt{Cache} stores a list of products indexed via unique keys with a prefix ''products:``. When users search for a specific product, the application verifies whether the search result has been cached. If the list of products is found in the cache (i.e., a ``hit''), the application returns the cached content. Otherwise (i.e., a ``miss''), it computes the list of products corresponding to the search query by accessing the high-latency component \texttt{ProductDB} (e.g., a large distributed database). The result of the slow computation is then cached to speed up future queries.

\begin{lstlisting}[caption={A code snippet of an e-commerce application to manage and retrieve products using a cache. This caching logic requires to change the control flow to deal with caching (line 7) and adds a key assignment (line 3) and its reference (lines 10 and 19). Finally, there are commands to remove content from the cache (line 19) to maintain consistency between source and cache.},label={lst:snippet1}]
public class ProductService {
 Cache cache = Cache.getInstance("productsCache");
 
 public List<Product> search(String query) {
  List<Product> products = cache.get(cachePrefix + query);
  if (products == null) {
   products = ProductDB.search(query);
   // add result to cache, expire in 30 seconds
   cache.put("products:" + query, products, 30);
  }
  return products;
 }
 
 public void update(Product product) {
  ProductDB.update(product);
  // ensuring consistency between cache and source
  // removing all cached queries starting with prefix
  cache.delete("products:*");
 }
}
\end{lstlisting}

\section{Benefits of Application-level Caching}
\label{sec:benefits}

When an application is able to reuse cached data, this leads to at least three benefits, listed as follows.

\begin{enumerate}
\item \textbf{Increased application throughput}: Avoiding long-running computations by using the previously computed results on the same input can yield a significant improvement in the execution time of the application because the application is able to process more requests per time unit. The performance improvement is higher if the cached computation is frequently requested.

\item \textbf{Reduced user perceived latency}: When a resource is cached, there is no need for downloading it from the network or querying a database. Even if a query is optimized, querying the database still requires network transport and a possibly expensive database-level computation. Avoiding these steps improves the application latency and makes the application more responsive to the end user. In addition, it enables the application to cope and perform well under an increased or expanding workload.

\item \textbf{Reduced application infrastructure costs}: An in-memory cache can serve a large number of requests to the same content, potentially reducing the need for database replicas or application server instances, thus reducing the overall application infrastructure costs. This is especially important in modern applications running in the cloud, where allocated resources are charged per processing time or throughput.
\end{enumerate}

Despite the ability to boost application's performance and throughput, application-level caching may sometimes cause performance degradation if the caching configuration is not designed and implemented correctly. A \emph{caching configuration} is the set of all settings and decisions to conceive and deploy application-level caching, including the selected cacheable content, cache size, and replacement policy.

To achieve the potential performance benefits, the caching configuration should consider the \emph{hit/miss ratio}, the \emph{characteristics of the cached data and workload}, and the \emph{frequency of data updates}. In the case of Listing~\ref{lst:snippet1}, the cache is likely to improve overall performance when the query is most of the time resolved via the \texttt{productsCache}.

The benefits of hit and the cost of a miss are dependent on the characteristics of the cached data and workload. For example, if calls for \texttt{ProductDB.search(query)} take a long time or are expensive to be computed, a small number of hits still means a significant improvement. In contrast, if calls for \texttt{ProductDB.search(query)} are almost as fast as performing a cache lookup, having a small number of misses does not necessarily imply significant improvements. In the same scenario, although the benefits of a single hit are minimal, if there is a huge number of calls in a short period, it may result in a significant improvement.

Finally, if a data update occurs, the cached content should be invalidated, and the next cache lookup will result in a miss. When a miss happens, data is loaded from the source and put in the cache again. If all these cache-related steps occur frequently, maintaining a cache may become even more expensive than simply querying the data directly from the database. Thus, to maximize the application performance, the caching configuration should consider how often the cached data gets updated.

\section{Costs of Application-level Caching}
\label{sec:costs}

The discussed benefits are the key motivation to adopt application-level caching. However, the example presented in Listing~\ref{lst:snippet1} also highlights that the benefits of application-level caching, presented in Section~\ref{sec:benefits}, come at the cost of dealing with non-trivial challenges from a software engineering point of view, as follows.

\begin{enumerate}
\item {\textbf{Code Maintainability.}} The current methodology for implementing application-level caching makes caching a cross-cutting concern, typically tangled with the application business logic. An additional long-term cost is increased maintenance efforts due to the intertwining of the caching logic with the application logic. In addition, cache keys are usually string-based and developers must pay extreme care because strings are hard to keep consistent with the existing code, as they usually are not handled by refactoring tools.

\item {\textbf{Understanding of the Application Behavior.}} The improvements that are achieved with caching depend on the characteristics of the application workload, which in turn depends on how many and how the users are typically accessing the application. For example, considering the expiration time of 30 seconds in Listing~\ref{lst:snippet1} (line 10), cached searches may not remain cached long enough to be reused if the users are performing different searches all the time.

\end{enumerate}

The specification of an appropriate cache configuration is difficult because workload patterns often change over time. Thus, not only may caching compromise code maintainability, but also must the caching configuration be continuously updated. For example, assume that now \texttt{ProductService} has a wider range of product categories because the company selling the products expanded its operations. This change potentially causes a significant increase in the number of users and, consequently, of searches for products, requiring developers to revise the cache configuration for these new conditions.

Although there are solutions to simplify the implementation of caches~\cite{Mertz2018a}, constantly adjusting the cache configuration and its implementation may be difficult in practice, in particular for large and complex applications~\cite{Chen2016}, because it involves: (i) detecting caching opportunities that can provide performance improvements, such as a recurrent computation that produce the same outputs based on the same inputs --- an activity that is not always trivial~\cite{Mertz2017}, (ii) managing consistency to achieve a better trade-off between performance and freshness of content, and (iii) understanding and dealing with possible inconsistencies caused by the caching logic. 

Because manually revising a caching configuration is time-consuming and error-prone, there is a need for methods and techniques that are sensitive to changing application workloads, to support developers in maintaining application-level caching. These approaches potentially extend the lifetime of a cache-related design, providing a long-term benefit and demanding less effort from developers~\cite{Mertz2018a}.


\section{Supporting Tools and Techniques}

\begin{figure}
\centering
\includegraphics[width=\linewidth]{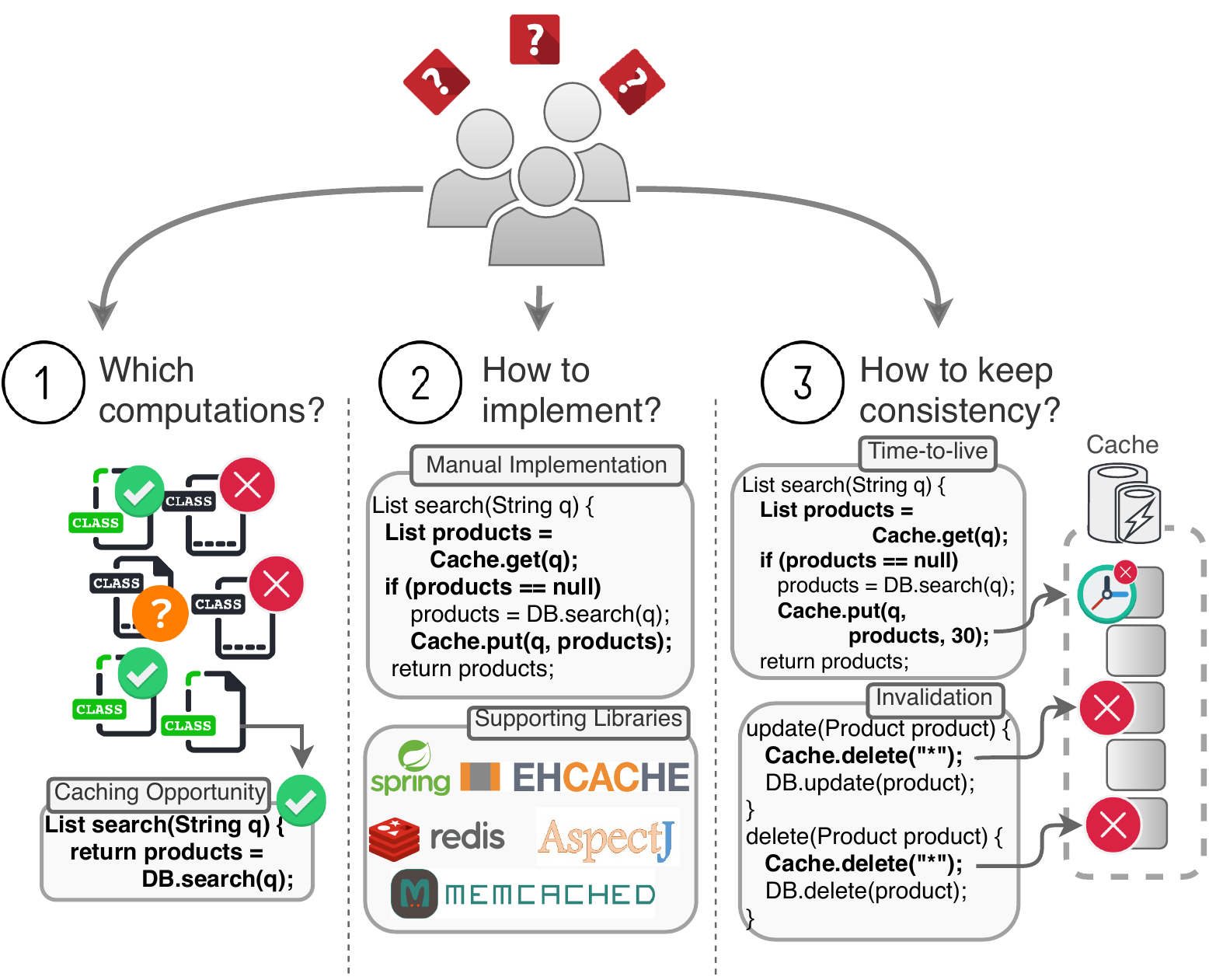}
\caption{Developer tasks for using application-level caching. To adopt application-level caching, the three main tasks to be performed by developers are: (1) choosing which computations to cache, (2) implementing caching decisions as a caching logic, potentially with the help of supporting libraries, and (3) defining a consistency strategy to keep the cached data consistent with their source.\label{fig:problems}}
\end{figure}

Designing and implementing a caching configuration requires developers to add caching logic to the code and deeply understand the application's behavior. Thus, developers must perform three main tasks, illustrated in Figure~\ref{fig:problems}, to adopt application-level caching: (1) choosing which computations to cache, (2) implementing a cache (and deciding on other caching configuration parameters, such as its size), and (3) keeping a cache consistent. As these are non-trivial tasks, recent tools and techniques have been developed, by both researchers and practitioners, to support the adoption of this type of caching. We next overview the types of support that exist to perform these tasks. A reader interested in deeply understanding individual approaches available in the literature is referred to elsewhere~\cite{Mertz2018a}.

\subsection{Which Computations to Cache?}

Application components with sub-optimal performance are typically the target of an analysis to decide what data to cache and consequently, to \emph{admit} cacheable content. The identification of such components is often manually performed by developers. To reduce the developers' effort in deciding what to cache and admitting cacheable content, many approaches have been proposed. Some of them make \emph{recommendations} of what to cache~\cite{Chen2018,DellaToffola2015}, while others provide \emph{automated support}~\cite{Mertz2018b,Chen2016}.

Recommender approaches target the application logic code that may execute expensive routines that repeatedly compute the same results. These approaches rely on profiling techniques to identify code locations where caching can be (safely) introduced and provide recommendations when the application may benefit from it (e.g., computing the same result multiple times, or creating similar data-structures). However, they demand a posterior manual analysis and implementation of such recommendations. Recommender approaches can be evaluated from different perspectives. The evaluation of Speedoo~\cite{Chen2018} assessed whether the recommendations were confirmed by developers by looking at multiple project stages, and the results show that 18.52\%--34.62\% of the recommended methods indeed undertook performance optimization during software evolution. MemoizeIt~\cite{DellaToffola2015} was evaluated by manually implementing the recommendations in existing applications and executing them against benchmarks to assess the provided improvements. It led to speedups by factors between 1.04x and 12.93x.

In contrast, automated approaches focus on automatically identifying and applying caching configurations, by placing a cache component as a tier within the application. These approaches target different granularity levels, e.g., method calls, database queries, data files, and web pages. This group of approaches removes the burden from developers to decide what to cache at the cost of losing the fine-grained control sometimes required by performance tuning. From the automated approaches, two approaches~\cite{Mertz2018b,Chen2016} change the application logic by applying caching to method calls in particular locations, thus avoiding to manually write caching code, such as in Listing~\ref{lst:snippet1}. Automated approaches are usually evaluated in terms of improvements in the throughput. APLCache~\cite{Mertz2018b} reported improvements from 2.78\% to 17.18\%. It also reported a hit/miss ratio change from -6.90\% to +36.13\% and was able to identify more cacheable opportunities in comparison with human-made decisions, i.e.,\ a caching manually developed. CacheOptimizer~\cite{Chen2016} demonstrated improvements in throughput by 27\%--138\%. It also automatically added caching to 6\%--55\% of the total number of possible cache locations.

A qualitative comparison of these two introduced groups of approaches is summarized in Table~\ref{tab:proscons}. Although automated approaches reduce the effort required to develop an application-level cache more than recommendation approaches, they are usually focused on specific content, which restricts the exploitation of certain caching opportunities. Consequently, automated approaches may result in lower performance improvements than guided labor-expensive solutions that are able to consider all application specificities and characteristics. In addition, automated approaches may demand a computing cost to acquire data at runtime and achieve an adequate caching configuration. In summary, the cost of automation may lower the performance gains.

Due to the differences in the evaluation settings adopted by existing automated caching approaches, i.e.,\ metrics, workload, applications, and granularity of cached data, the reported evaluation results are not comparable. Empirical studies to directly compare the benefits of existing approaches will be valuable future work. Finally, existing approaches in both groups are implemented as prototypes, used to evaluate their effectiveness, but further work is needed to make them solutions to be used in industrial settings.

\begin{table}
\footnotesize
\begin{tabular}{p{4cm}p{4cm}}
\toprule
\textbf{Recommendation approaches} & \textbf{Automated approaches} \\ \midrule
(+) Summarized actionable recommendation & (-) Limited transparency regarding caching decisions \\
(+) Guidance for optimization implementation & (-) Imply overhead \\
(+) Applied to general content & (-) Restricted to content-specific contexts \\
(-) Demand refactoring & (+) Seamless integration \\
(-) Reduced maintainability & (+) Improved separation of concerns \\
(-) Need to be reexecuted from time to time & (+) Capable of adapt itself to the application workload changes \\
(-) Error-prone implementation & (-) Susceptible to false positives \\ \bottomrule
\end{tabular}
\caption{Pros and cons of approaches that support the identification and the automated management of cacheable computations. \label{tab:proscons}}
\end{table}

\subsection{How to Implement Caching?}

Because application-level caching is a cross-cutting concern, in addition to its design, its implementation is also a challenge. Fortunately, there are several mature techniques to ease this task.

There are approaches that assist developers by providing consolidated knowledge regarding the design and implementation of application-level caching, thus reducing the effort while implementing caching. These approaches include supporting libraries and frameworks, such as \emph{Caffeine} and \emph{Spring Caching}, which provide developers with customizable and reusable APIs that ease the manipulation of the cache system and its content, improving software maintainability. For example, Listing~\ref{lst:snippet2} shows the implementation of the same functionality as in Listing~\ref{lst:snippet1}, but using Spring Caching to handle the cache component. The use of annotations (lines 1, 4 and 11) reduces the amount of code needed to manipulate the cache and removes the caching code from the method body, thus reducing the coupling between the caching logic and the application base code. However, it does not support to cache partial results within methods, which may demand code refactoring in some situations.

Beyond concrete libraries and frameworks that support developers in implementing caching, there are guidelines and patterns~\cite{Mertz2017} derived from the analysis of the state of the practice. These patterns are similar to software design patterns, as they provide reusable abstract solutions to a recurring problem within a given context in software design. Although patterns demand implementation effort and code refactoring to be put in practice, they improve code maintainability by guiding developers to a reproducible solution.

\begin{lstlisting}[caption={\emph{Annotation-based Caching}. Implementation of methods similar to those in Listing~\ref{lst:snippet1} using Spring Caching to handle cacheable content.},label={lst:snippet2}]
@CacheConfig(cacheNames="productsCache")
public class ProductService {
  
  @Cacheable(key="'products:' + #query", ttl = 30,
             unless = "#result != null")
  public List<Product> search(String query) {
    List<Product> products = ProductDB.search(query);
    return products;
   }
   
  @CacheEvict("products:*")
  public void update(Product product) {
    ProductDB.update(product);
  }
}
\end{lstlisting}

Caching implementation can be also addressed by reusable building blocks which are based on declarative domain-specific languages~\cite{Gupta2011,Ports2010}. From an implementation perspective, they provide a solution similar to that presented in Listing~\ref{lst:snippet2}, where developers simply describe what to cache via annotations processed at compilation time. In addition, these approaches automatically and proactively manage consistency by tracking changes in the data source at runtime and, consequently, they do not require specific descriptions of how and when to evict cached content. Although they are less customizable and flexible than the annotations presented in Listing~\ref{lst:snippet2}, these approaches provide an effortless and automated solution to caching.

Finally, seamless solutions do not demand any cache-related implementation, which is especially useful in situations where refactoring an application to add caching is practically infeasible. As caching is a cross-cutting concern, aspect-oriented programming (AOP) or similar techniques have been explored towards increasing the separation of concerns and modularity of application-level caching~\cite{Surapaneni2011,Bouchenak2006}. Despite being effortless and automated, the flexibility and customizability of these approaches are restricted.

\subsection{How to Keep Caches Consistent?}

Because caching leads to duplicated data, maintaining consistency between the original and cached data becomes an important concern. In manually implemented application-level caches, one of the commonly adopted practices is a time-based expiration policy. This involves setting a fixed and arbitrary timeout to each item in the cache, which indicates when it becomes invalid, such as presented in Listings~\ref{lst:snippet1} and \ref{lst:snippet2} (lines 10 and 4, respectively). However, this strategy compromises data freshness, mainly when the same timeout to purge stale data is used for all the items in the cache. A short expiration time gives higher consistency but tends to result in lower hit ratios; while a long expiration time can provide higher availability at the cost of occasionally returning stale data. 

To solve this problem of defining an adequate expiration time, called \emph{time-to-live} (TTL), some approaches~\cite{Huang2010} propose using machine learning models or utility functions to define dynamic TTL values based on how users access content or the logs of database queries. These approaches provide increased adaptiveness and seamlessness for developers, which are freed of defining hard-coded expiration values. However, they lack of customizability, and may not be adequate for all application domains and workloads, possibly resulting in stale content.

An alternative to expiration-based consistency is \emph{invalidation}, in which cached content is evicted when its source data changes. No stale data is therefore returned. This requires to track changes in the source of information, trace it to the cached content, and finally trigger invalidation actions. Many approaches that cache database content rely on invalidation-based consistency~\cite{Ports2010,Bouchenak2006}. These approaches monitor database queries to detect data changes and its dependencies to evict cached content when write operations are performed, which can be done through a middleware acting as a bridge between the database and the application, or even by inserting triggers directly on the database code. Some approaches use manual input from developers, who should indicate cacheable functions or provide a general description of cacheable content with its dependencies and other metadata~\cite{Ports2010}. Despite the manual input, invalidation-based approaches provide a simple and effortless solution to ensure data freshness. Other approaches go further and automatically analyze the data dependency graph or query statements to identify cached content related to the recently updated data~\cite{Bouchenak2006}, towards providing a seamless and automatic invalidation.

\section{Main Open Challenges}

Looking beyond existing tools and techniques that facilitate the adoption of application-level caching, several challenges remain open. We next outline these challenges, hoping to inspire future work that addresses them.

\subsection{Adaptive Caching Configurations}

To provide effective caching configurations, an in-depth understanding of the application behavior and usage patterns is required. However, as the application evolves and so its workload, caching configurations may (soon) become ineffective. Keeping an updated caching configuration can be cumbersome because it demands time and effort from developers that must periodically review their caching decisions. Moreover, there is no guarantee that a pre-computed caching decision taken at development time will be effective once the application is deployed.

This situation calls for adaptive approaches that tailor caching configurations to the current workload. This can be done, in a simpler way, with pre-defined caching configurations that are activated under certain conditions of the application behavior. Alternatively, more sophisticated approaches can be proposed to automate caching decisions at runtime. Providing adaptive and automated caching configurations is possible but imposes two challenges discussed next.

\subsection{Cost-Effective Runtime Monitoring}

Adaptive approaches that select suitable caching configurations on the fly require updated information about the application to understand the current workload. Monitoring techniques that capture runtime data can support adequate decisions in a timely fashion. However, capturing runtime data of the application behavior during its execution might itself have a negative impact on the application's performance, possibly significantly reducing or even cancelling the gains obtained by caching.

To reduce the overhead of collecting dynamic information during an execution, knowledge collected with static and offline analyses of the program code can be leveraged. This can be used, for example, to identify candidates of pre-computed caching configurations, or to remove the number of probes inserted to collect data. Alternatively, a static analysis, possibly supported by developer-provided annotations, could identify different phases in the execution, in which the workload changes from one phase to another. Based on these phases, the analysis could infer how to modify the cache content when transitioning from one phase to another.

\subsection{Prediction of Effective Caching Configurations}

A common problem of caching, no matter whether at the application level or not, is that a cache usually lacks behind in its decision: the cache must first be warmed up before becoming effective and it only learns which data to evict by means of cache misses or a timeout. Instead, application-level caching could benefit from adapting a well-known idea from hardware-level caching: to learn and predict what data to pre-compute (prefetching) or to evict. Despite the existence of these kind of approaches, which make use of machine learning or other sophisticated techniques, in other caching layers and applications (such as search engines and proxy level), at the application level there is a lack of practical solutions in this direction, given the arbitrary data an application can manipulate. For example, a runtime profiler could learn pre-computation and eviction policies from representative, pre-deployment executions, which can then be used to speed-up future executions of the deployed application.

\subsection{Correctness Assurance}

One requirement for cacheable computations is that a computation repeatedly produces the same output based on the same input. In addition, two requirements are crucial to guarantee that caching does not compromise the correctness of the application. First, any code location that changes the input data of the cached computation must be identified to invalidate the cached content and avoid stale data. Alternatively, as discussed, a TTL can be used. Second, the cached computation must be carefully inspected to ensure that it does not have any desired side effects that would be suppressed by caching. Examples of computations with such side effects are methods that change the state of objects or perform globally visible operations. Consequently, it is necessary to employ techniques that not only identify cacheable spots but also identify situations where caching should not be performed. This can also be tackled by using static analyses of the source code or developer-provided indications of non-cacheable operations.

\section{Conclusion}

In summary, substantial effort has been made to support the development of application-level caching, which can save precious computing cycles by reusing already computed values. However, its implementation and maintenance are non-trivial and there are many open challenges that must be addressed to further reduce the manual effort required for adopting application-level caching.

\bibliographystyle{IEEEtran}
\bibliography{references}

\renewenvironment{IEEEbiography}[1]
  {\IEEEbiographynophoto{#1}}
  {\endIEEEbiographynophoto}

\vskip -2\baselineskip plus -1fil

\begin{IEEEbiography}{Jhonny Mertz}
is a software engineer at HP Inc. and Ph.D. student at Universidade Federal do Rio Grande do Sul (UFRGS), Brazil. His research interests are in the intersection of software engineering and data analytics, particularly interested in leveraging software traces and logs to support software development and operations.
\end{IEEEbiography}

\vskip -2\baselineskip plus -1fil

\begin{IEEEbiography}{Ingrid Nunes}
is a senior lecturer at Universidade Federal do Rio Grande do Sul (UFRGS), Porto Alegre, Brazil, since 2012. Her main research interests are software maintenance and evolution and self-adaptive systems. She completed her phd at the Pontifical Catholic University of Rio de Janeiro (PUC-Rio) in 2012, with a one-year sandwich period at King's College London, and was on a one-year sabbatical year at TU Dortmund, with a CAPES-Alexander von Humbolt Fellowship in 2016-2017.
\end{IEEEbiography}

\vskip -2\baselineskip plus -1fil

\begin{IEEEbiography}{Luca Della Toffola}
is a software engineer at xorlab AG. Previously he was at ETH Zurich, where he received his PhD in 2018. His research focused on program analysis and program synthesis for performance anomalies detection.
\end{IEEEbiography}

\vskip -2\baselineskip plus -1fil

\begin{IEEEbiography}{Marija Selakovic}
is a research engineer at Huawei Research Center in Dresden. Previously, she was at TU Darmstadt, where she received her Ph.D. in 2019. Her research interests include program analysis and automated test generation for improving software quality. Marija also has a master's degree in computer science from VU University Amsterdam and the University of L'Aquila and a bachelor's degree in information systems from the University of Belgrade.
\end{IEEEbiography}

\vskip -2\baselineskip plus -1fil

\begin{IEEEbiography}{Michael Pradel}
is a full professor at University of Stuttgart, which he joined after a PhD at ETH Zurich, a post-doc at UC Berkeley, an assistant professorship at TU Darmstadt, and a sabbatical at Facebook. His research interests span software engineering, programming languages, security, and machine learning, with a focus on tools and techniques for building reliable, efficient, and secure software.
\end{IEEEbiography}

\end{document}